\documentclass[onecolumn]{emulateapj}  
\usepackage{epstopdf}
\usepackage{ulem}
\usepackage{amssymb}
\usepackage{natbib}
\usepackage{times}
\usepackage{graphicx}
\usepackage[usenames,dvipsnames]{pstricks}
 \usepackage{psfrag}
                                                                                        
\usepackage[colorlinks=true,linkcolor=blue,citecolor=blue]{hyperref}

\def\apj{ApJ}%% Astrophysical Journal
\def\apjl{ApJ}%% Astrophysical Journal, Letters
\def\apjs{ApJS}%% Astrophysical Journal, Supplement
\def\jcap{J. Cosmology Astropart. Phys.}%% Journal of Cosmology and Astroparticle Physics
\def\mnras{MNRAS}%% Monthly Notices of the RAS
\def\prd{Phys.~Rev.~D}%% Physical Review D
\def\nat{Nature}%% Nature
\newcommand{\be}{\begin{equation}}
\newcommand{\ee}{\end{equation}}
\newcommand{\bary}{\begin{eqnarray}}
\newcommand{\eary}{\end{eqnarray}}
\newcommand{\en}{E_\nu}

\shorttitle{PeV Neutrino}
\shortauthors{Fraija et al.}
\begin{document}
\title{Could a multi-PeV neutrino event have as origin the internal shocks  inside the  GRB progenitor star?}
\author{N. Fraija\altaffilmark{1}}
\affil{Instituto de Astronom\'ia, UNAM, M\'exico, 04510}
\email{nifraija@astro.unam.mx}
\altaffiltext{1}{Instituto de Astronom\' ia, Universidad Nacional Aut\'onoma de M\'exico, Circuito Exterior, C.U., A. Postal 70-264, 04510 M\'exico D.F., M\'exico}
%
%\date{\today} % It is always \today, today,
\begin{abstract}
The IceCube Collaboration initially reported the detection of 37 extraterrestrial neutrinos in the TeV - PeV energy range. The reconstructed neutrino events were obtained during three consecutive years of data taking, from 2010 to 2013. Although these events have been discussed to have an extragalactic origin, they have not been correlated to any known source.   Recently, the IceCube Collaboration reported a neutrino-induced muon event with energy of $2.6\pm0.3$ PeV which corresponds to the highest event ever detected.  Neither the reconstructed direction of this event (J2000.0), detected on June 11 2014 at R.A.=110$^\circ$.34, Dec.=11$^\circ$.48  matches with any familiar source.    Long gamma-ray bursts (lGRBs) are usually associated with the core collapse of massive stars leading relativistic-collimated jets inside stars with high-energy neutrino production. These neutrinos  have been linked to the 37 events previously detected by IceCube experiment. In this work,  we explore the conditions and values of parameters  so that the highest neutrino recently detected could be generated by proton-photon and proton-hadron interactions at internal shocks inside lGRB progenitor star and then detected in IceCube experiment.  Considering that internal shocks take place in a relativistic collimated jet, whose (half) opening angle is $\theta_0\sim$ 0.1, we found that lGRBs with total luminosity $L\lesssim 10^{48}$ erg/s and internal shocks on the surface of progenitors such as Wolf-Rayet (WR) and blue super giant (BSG) stars  favor this  multi-PeV neutrino production, although this neutrino could be associated to $L\sim 10^{50.5}$ ($\sim 10^{50}$) erg/s  provided that the internal shocks occur at $\sim 10^9$ ($\sim 10^{10.2}$) cm for a WR (BSG).
%PACS numbers may be entered using the \verb+\pacs{#1}+ command.
\end{abstract}

%$10^{46.5}\leq L\leq 10^{50.5}$ erg/s ($10^{46}\leq L\leq 10^{49.8}$ erg/s) and  $10^8\leq r_j\leq 10^{11}$ cm ($10^{10}\leq r_j\leq 10^{12.3}$ cm), respectively for a WR (BSG) with $\Gamma\leq40$ ($\Gamma\leq220$).


%\pacs{98.70.Rz; 98.70.Sa}% PACS, the Physics and Astronomy
%\maketitle

%Fanaroff-Riley Class I (FRI)
\section{Introduction}\label{sec-Intro}
\begin{figure*}
\centering
\includegraphics[width=0.7\textwidth]{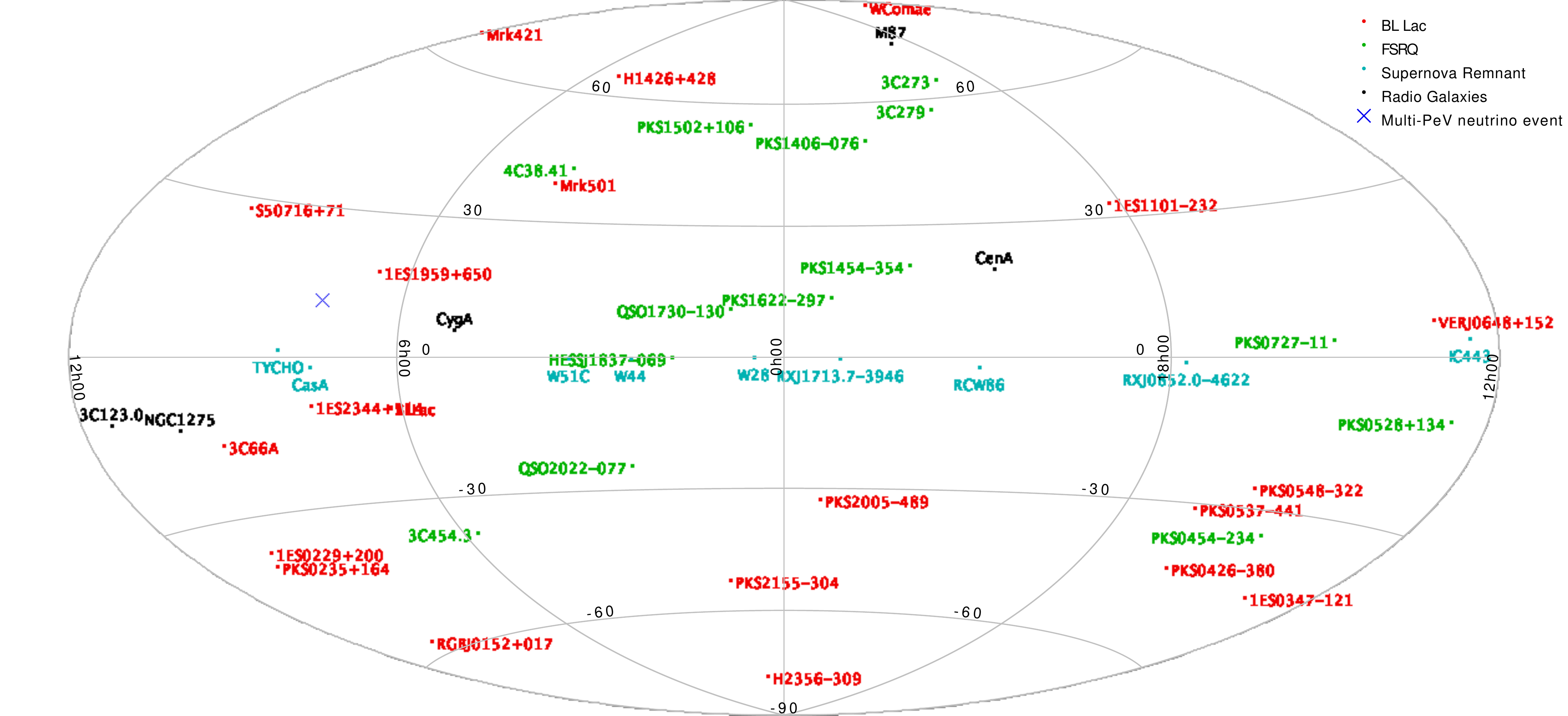}
\caption{Sky-map with the location of BL Lac sources, FSRQs, Supernova Remnants, Radio Galaxies and the multi-PeV neutrino event  detected by the IceCube detector (blue cross). \label{skymap}}
\end{figure*}

%
%%%%%%%%%%%%%%%%%%%%%%%%%%%%%%%%%%%%%%%%%%%%%%%%%%%%%%%%%%%%%%%%%%%%%%%%%%%%%%%%%%%%%%%%%%
%%%%%%%%%%%%%%%%%%%%%%%%%    Neutrino                               %%%%%%%%%%%%%%%%%%%%%%%%%%%%%%%%%%%%%%%%%%%%%%%%%%
%%%%%%%%%%%%%%%%%%%%%%%%%%%%%%%%%%%%%%%%%%%%%%%%%%%%%%%%%%%%%%%%%%%%%%%%%%%%%%%%%%%%%%%%%%
%
The IceCube Collaboration has reported evidence for extragalactic neutrinos. They announced, firstly,  two shower-like events of PeV neutrinos \citep{2013PhRvL.111b1103A}, after  24 events (19 shower-like, and the remainder track-like; \cite{2013Sci...342E...1I})  and finally 11 events  were added to these detections \citep{2014PhRvL.113j1101A}. In the aggregate, 37 events (9 track-like and 28 shower-like) were detected after analyzing three year data (2010 to 2013; \cite{2014PhRvL.113j1101A}).  The excess measured over the atmospheric background had a statistical significance 5.7$\sigma$. Arrival directions of these neutrino events seem to be isotropic and nonneutrino track-like has been associated with the location of known sources yet.
Recently, a multi-PeV neutrino-induced muon event detected in  2014 June 11 was reported by the IceCube Collaboration \citep{2015ATel.7856....1S}. The neutrino with an energy $2.6\pm0.3$ PeV is the highest-energy event ever detected.  This event had a probability of less than 0.01\% of having an atmospheric origin, hence putting together in a skymap this event with the known sources as shown in Figure \ref{skymap}, the reconstructed direction of this event (J2000.0: R.A.=110$^\circ$.34, Dec.=11$^\circ$.48) does not match with any of the known sources.\\
%
%%%%%%%%%%%%%%%%%%%%%%%%%%%%%%%%%%%%%%%%%%%%%%%%%%%%%%%%%%%%%%%%%%%%%%%%%%%%%%%%%%%%%%%%%%
%%%%%%%%%%%%%%%%%%%%%%%%%%     Core Collapse                               %%%%%%%%%%%%%%%%%%%%%%%%%%%%%%%%%%%%%%%%%%%%%%
%%%%%%%%%%%%%%%%%%%%%%%%%%%%%%%%%%%%%%%%%%%%%%%%%%%%%%%%%%%%%%%%%%%%%%%%%%%%%%%%%%%%%%%%%%
%
Long gamma-ray bursts (lGRBs) have been generally linked to core collapse of massive stars leading to relativistic-collimated jets and supernovae (CCSNe) of type Ib,c and II. Type II and Ib are widely believed  to have a radius of $R_\star\approx 3\times 10^{12}$ cm, usually associated with blue super giant (BSG) stars, and type Ic supernovae are thought to be Wolf-Rayet (WR) stars with radius $R_\star\approx $ 10$^{11}$ cm.  Depending on the luminosities and durations, successful lGRBs have revealed a variety of GRB populations: low-luminosity (ll),  ultra-long (ul) and high-luminosity (hl) GRBs \citep{PhysRevLett.87.171102,0004-637X-662-2-1111, 0004-637X-766-1-30}. While llGRBs and ulGRBs have a typical duration of ($\sim$ 10$^3$  s), hlGRBs have a duration of tens to hundreds of seconds. Another relevant population associated with CCSNe, although unobservables in photons, are failed  GRBs which could be much more frequent than successful ones, limited only by the ratio of type Ib/c and type II SNe to GRBs rates.  This population has been  identified by having high-luminosities, mildly relativistic jets and durations from some to ten seconds \citep{2002MNRAS.332..735H, PhysRevLett.87.171102,2006Natur.442.1014S, 2010Natur.463..513S}.\\
%
%%%%%%%%%%%%%%%%%%%%%%%%%%%%%%%%%%%%%%%%%%%%%%%%%%%%%%%%%%%%%%%%%%%%%%%%%%%%%%%%%%%%%%%%%%
%%%%%%%%%%%%%%%%%%%%%%%%%%     What is discussed in literature in this environment %%%%%%%%%%%%%%%%%%%%%%%%%%%%%%%%%%%%%
%%%%%%%%%%%%%%%%%%%%%%%%%%%%%%%%%%%%%%%%%%%%%%%%%%%%%%%%%%%%%%%%%%%%%%%%%%%%%%%%%%%%%%%%%%
%
%
Many astrophysical environments have been explored to explain these high-energy (HE) neutrinos. For instance, GRB \citep{2013PhRvL.111l1102M, 2013PhRvD..88j3003R, 2014MNRAS.437.2187F, 2013ApJ...766...73L, 2015JCAP...09..036T, 2014MNRAS.445..570P, 2015arXiv151201559T}, active galactic nuclei (AGN; \cite{2014JHEAp...3...29D,2013PhRvD..88d7301S, 2015MNRAS.448.2412P, 2015MNRAS.452.1877P}), dark matter (DM; \cite{2013PhRvD..88a5004F, 2013JCAP...11..054E}) and cosmic neutrino background connections \citep{2014PTEP.2014f1E01I, 2014PhRvD..90f5035N}. \\ 
GRBs have been widely pointed out  as emitter sources of HE neutrinos in connection to  cosmic rays accelerated up to ultra high energies  \citep{1995ApJ...449L..37M, 1995ApJ...452L...1W, 1995PhRvL..75..386W, 1999PhRvD..59b3002W, 2000NuPhS..87..345W, 2000ApJS..127..519W, 1997PhRvL..78.4328V, 1998PhRvD..58l3005R, 1998ApJ...499L.131B}.\\    
Recently, \cite{2013PhRvL.111l1102M} studied HE neutrino production in collimated jets inside progenitors of lGRBs, considering both collimation and internal shocks. Authors showed that whereas classical GRBs may be too powerful to generate HE neutrinos inside stars,  low-power (lp) GRBs (including llGRBs) were candidates for producing the HE neutrinos reported by the IceCube collaboration.\\
One major progress in recent years has been the realization that photon emission from the photosphere, namely optically thick regions may be relevant to GRBs. Clearly, the photosphere marks the point below which the plasma is optically thick and above which it is optically thin. Several theoretical works in recent years dealt with the properties of shock waves that propagate in optically thick regions - the so called "radiation mediated shocks" \citep{2010ApJ...725...63B, 2010ApJ...716..781K, 2011ApJ...733...85B, 2012ApJ...756..174L}.  These authors have argued, that the properties of these radiation mediated shocks in the depth of the star  are such that do not allow them to accelerate protons up to HEs.\\
In this work, we explore the conditions under which the highest neutrino recently detected at 2.6 PeV could be created  by hadronic interactions in internal shocks inside a GRB progenitor star.  In section II, we show the dynamics of jet, internal shocks, particle acceleration process and the neutrino production by proton-photon (p-$\gamma$) and proton-hadron (p-h) interactions.  In section III, we explain the neutrino flux expected on IceCube detector  from p-$\gamma$ and p-h interactions.   In section IV we discuss our results and  in section V we give a brief conclusions. We hereafter use primes (unprimes) to define the quantities in a comoving (observer) frame, $Q_x\equiv Q/10^x$ in c.g.s. units and $k=\hbar$=c=1 in natural units.
\section{Internal shocks and Neutrino Production}
In GRB context,  one of the most prosperous theory to explain the prompt emission and the afterglow is  the fireball model \citep{2004IJMPA..19.2385Z,2006RPPh...69.2259M}.  In this model,  the prompt emission is explained through internal shell collisions, where faster shells ($\Gamma_f$) catch slower  shells ($\Gamma_s$).  These internal shocks take place at a distance of ${\small r_j=2\Gamma^2\,t_\nu}$, with $t_v$ given by  the variability time scale of the central object  and  $\Gamma\simeq\sqrt{\Gamma_f\,\Gamma_s}$ the bulk Lorentz factor of the propagating shock.   We are interested in those internal shocks that occur inside the star  (r$_j< R_*$ with $R_*$ the radius of the progenitor's stellar surface) and the hydrodynamic jet is collimated by the cocoon pressure via collimation shocks \citep{2011ApJ...739L..55B, 2013ApJ...777..162M}.   It happens when the jet luminosity is low and/or density is high for an initial opening angle.   The total energy density $U=1/(8\,\pi\,m_p)\,\Gamma^{-4}\,L\,t^{-2}_{\nu}$ in internal shocks is equipartitioned to amplify the magnetic field
\be
B'= \epsilon_B^{1/2}\,\Gamma^{-2}\,L^{1/2}\,t^{-1}_\nu\,,
\label{mfield}
\ee
with $\epsilon_B=U_B/U=(B^2/8\pi)U$ and accelerate particles $\epsilon_i=U_i/U$, where {\rm i} is the {\rm ith} kind of particle with the condition $\epsilon_B + \sum_i \epsilon_i =1$.  Electrons are accelerated in internal shocks and then are cooled down rapidly by synchrotron radiation in the presence of the magnetic field. The synchrotron photons in this length scale produce an opacity to Thomson scattering given by {\small $\tau_{th}'=\frac{\sigma_T}{4\pi\,m_p} \Gamma^{-3}\,L\,t^{-1}_\nu$} where $\sigma_T$ is the Thompson cross section and m$_p$ is the proton mass.
Photons can thermalize in a black body temperature with peak energy given by
\be\label{enph}
T'_{\gamma}\simeq\frac{1.2}{\pi}\,\epsilon_e^{1/4}\, L^{1/4}\,\Gamma^{-1}\,t_v^{-1/2}.
\ee
The proton distribution is cooled down in internal shocks  via electromagnetic (synchrotron radiation and inverse Compton (IC) scattering) and hadronic  p-$\gamma$ and p-h interactions channels.  P-$\gamma$ and p-h interactions take place when accelerated protons interact  with  thermal keV  photons and proton density at the shock.   In both interactions HE charged pions and kaons are produced;  $p+\gamma/h \to X+\pi^{\pm}/K^{\pm}$, and subsequently HE neutrinos  $\pi^+\to \mu^++\nu_\mu\to e^++\nu_e+\bar{\nu}_\mu+\nu_\mu$ and $\pi^-\to \mu^-+\bar{\nu}_\mu\to e^-+\bar{\nu}_e+\nu_\mu+\bar{\nu}_\mu$. \\
\subsection {Particle acceleration process}
Before the jet breaks out the star, it keeps the bulk Lorentz factor low by converging cylindrically via collimation shocks under the cocoon pressure.  The characteristic radius to which the shocked jet becomes cylindrical is \citep{2011ApJ...739L..55B}
{\small
\be\label{rc}
r_c=\left[\frac{3\varepsilon_c^2}{16\pi^{3/2}\eta_c\chi_c}\right]^{1/5}\,t^{2/5}\,L^{3/10}_j\,\theta^{4/5}_0\,\rho_a^{-3/10}\,,
\ee
}
where $\rho_a$ is the ambient density,  $\theta_0$ is initial opening angle,  $\varepsilon_c=\frac{5}{3+\alpha_\delta}$, $\eta_c=\frac{3}{3-\alpha_\delta}$ and $\chi_c=\frac{5}{7-\alpha_\delta}$.  Here, $L_j=\theta^2_0\,\frac{L}{4}$ is the absolute jet luminosity defined through the total luminosity L.   In the meantime, the position of the non-relativistic head is 
{\small
\be\label{rh}
r_h=\left[\frac{16\eta_h\zeta^2_h}{3\pi}\right]^{1/5}\,t^{3/5}\,L^{1/5}_j\,\theta^{-4/5}_0\,\rho_a^{-1/5}\,,
\ee
}
where  $\eta_h=\frac{3}{3-\alpha_\delta}$ and $\zeta_h=\frac{5-\alpha_\delta}{3}$.  We consider density profiles as
{\small
\bary\label{rhoa}
\rho_a \simeq
\cases{ 
\frac{(3-\alpha)M_\star}{4\pi R^3_\star } \left(\frac{r}{R_\star}\right)^{-\alpha_\delta}\,\,{\rm for\, WR} \, \cr
\rho_0\,  K\,\hspace{2cm}\,{\rm for\, BSG}\,, 
 } 
\eary
}
\noindent with  $K$ given by $\left(\frac{R_\star}{r}\right)^{17/7}$ for $r_a< r < r_b$ and $\left(\frac{R_{\star}}{r}\right)^{17/7}\frac{\left(r-R_{\star}\right)^{5}}{\left(r_{b}-R_{\star}\right)^{5}}$ for $r>r_b$. Here $\rho_0=3.4\times10^{-5}\,\mathrm{g\,cm^{-3}}$, $r_{a}=10^{10.8}$ cm, $r_{b}=10^{12}$ cm, $M_\star$ is the progenitor mass and $\alpha_\delta$ = 3/2 - 3, being 3/2 and 3 for convective and radiative envelopes, respectively.
From eqs. (\ref{rc}), (\ref{rh}) and (\ref{rhoa}),  the cylindrical and head radii for a WR are
{\small
\bary\label{wr}
r_c&=&\left[\frac{3\varepsilon^2_c}{16(3-\alpha_\delta)^{3/2}\eta_c\chi_c} \right]^{\frac{2}{10-3\alpha_\delta}}\, t^\frac{4}{10-3\alpha_\delta}\,L^\frac{3}{10-3\alpha_\delta}\theta^\frac{14}{10-3\alpha_\delta}_0\,M^\frac{3}{3\alpha_\delta-10}_\star\,R^\frac{9-3\alpha_\delta}{10 - 3\alpha_\delta}_\star \cr
r_{h}&=&\left[\frac{16 \eta_h\zeta^2_h}{3(3-\alpha_\delta)} \right]^{\frac{1}{5-\alpha_\delta}}\, t^\frac{3}{5-\alpha_\delta}\,L^\frac{1}{5-\alpha_\delta}\,\theta^{-\frac{2}{5-\alpha_\delta}}_0\,M^\frac{1}{\alpha_\delta-5}_\star\,R^\frac{3-\alpha_\delta}{5 - \alpha_\delta}_\star\,, 
\eary
}
and for a BSG are\\ 
{\small
\bary\label{bsg}
r_c&=&\left[\frac{3\varepsilon^2_c}{2^7\pi^{3/2}\eta_c\chi_c}\right]^{1/5}\,t^{2/5}\,L^{3/10}\,\theta^{7/5}_0\,\rho_0^{-3/10}\,K^{-3/10}\cr
r_h&=&\left[\frac{2^3\eta_h\zeta^2_h}{3\pi}\right]^{1/5}\,t^{3/5}\,L^{1/5}\,\theta^{-2/5}_0\,\rho_0^{-1/5}\,K^{-1/5}\,.
\eary
}
Propagating shocks in optically thick regions are  radiation mediated.  The properties of these radiation mediated shocks in the depth of the star  are such that do not allow to accelerate protons up to HEs.  Upstream proton flow is decelerated by photons created in the downstream and diffusing into the upstream region \citep{2010ApJ...725...63B, 2010ApJ...716..781K, 2011ApJ...733...85B, 2012ApJ...756..174L}. It happens when the deceleration scale  $l_{dec}\simeq (n_p\,\sigma_T y_{\pm})^{-1}$ of the shock width is smaller than the comoving size ($l_\nu$) of the upstream flow, as indicated by \cite{2013PhRvL.111l1102M}. Protons, then, are expected when shock propagates in an optically thin region (radiation unmediated shock; $l_\mu\ll l_{dec}$). Hence,  accelerated protons are expected around the photosphere $\tau\sim(1-10)$ \citep{2008PhRvD..78j1302M}, as assumed in the dissipative photosphere scenario \citep{ 2005ApJ...628..847R, 2006ApJ...651L...5M}.  Following \citet{2013PhRvL.111l1102M}, a reasonably necessary condition for proton acceleration is
\be\label{opt_depth}
\tau= n_p\,\sigma_T\,r\lesssim\,\, {\rm min}[\Gamma^2_{rel}, 0.1C^{-1}\Gamma^3_{rel} ]\,,
\ee
where $C=1+2\,{\rm ln}\,\Gamma^2_{rel}$,  $\Gamma_{rel}\approx(\Gamma_f/\Gamma+\Gamma/\Gamma_f)/2$ is the relative Lorentz factor between the faster and merged shells and $n_p$ is the number proton density.\\
In this case,   neutrino energy is limited by the  maximum energy at accelerated protons which is obtained equaling the  time scales of acceleration  ${\small t'_{acc}=\frac{2\pi\xi\,B'_{c,p}}{m_p^2}\,E'_p\,\epsilon^{-1/2}_B\, L^{-1/2}\,\Gamma^{2}\,t_v}$ and cooling synchrotron ${\small t'_{p,syn}=\frac{6\pi\,m_p^4}{\sigma_T\,m_e^2\,E'_p}\,\epsilon^{-1}_B\, L^{-1}\,\Gamma^{4}\,t^2_v}$, supposing that the synchrotron process is the longest timescale (e.g. see \cite{2014MNRAS.437.2187F}). Hence, it can be written as
\be\label{Epmax}
E_{p,max}=\biggl(\frac{3\,e\,m_p^4}{\sigma_T\,\xi\,m_e^2}\biggr)^{1/2}\,\epsilon^{-1/4}_B\, L^{-1/4}\,\Gamma^2\,t^{1/2}_v\,,
\ee
where e is the electric charge, m$_e$ is the electron mass, $\xi$ is a factor of equality and $B'_{c,p}$ is the critical magnetic field for protons.
\subsection {P-$\gamma$ interactions}
The number density of thermalized photons $ \eta'_\gamma\simeq \frac{2\zeta(3)}{\pi^2}\,T'_\gamma$ can be explicitly written as
\be\label{ngamma}
 n'_\gamma=\frac{3.2\,\zeta(3)}{\pi^5}\,  \epsilon_e^{3/4} L^{3/4}\, \Gamma^{-3} t_v^{-3/2}\,.
\ee
The optical depth characteristic for this process  is ${\small \tau'_{p\gamma}=\frac{6.5\,\zeta(3)\sigma_{p\gamma}}{\pi^5}\,  \epsilon_e^{3/4} L^{3/4}\, \Gamma^{-2} t_v^{-1/2}}$, where $\zeta(x)$ is the zeta function and $\sigma_{p\gamma}$ is the cross section of p-$\gamma$ interactions.   From the resonance $p+\gamma \to \Delta^+$ in the rest frame and taking into account that  each neutrino  carries 5\%  of the  proton energies ($E_\nu=1/20\,E_p$), then the threshold neutrino energy can be written as
\be
E_{\nu,\pi}\simeq 10^{-1.67}\,\pi\,(m^2_\Delta-m_p^2)\, \epsilon^{-1/4}_e\, L^{-1/4}\,\Gamma^2\,t^{1/2}_v,
\ee
where $m_\Delta$ is the resonance mass.
\subsection {P-h interactions}
Protons co-accelerated in the internal shocks have a comoving number density \citep{2010JHEP...03..031R, 2005MPLA...20.2351R, and05,2013MNRAS.432..857M}
\be\label{np}
n'_p=1/(8\,\pi\,m_p)\,\Gamma^{-4}\,L\,t^{-2}_{\nu}\,,
\ee
with energy density  $n'_p m_p$.  The  optical depth for  p-h  interactions is ${\small \tau'_{pp}\simeq\frac{\sigma_{pp}}{4\,\pi\,m_p}\,  L\, \Gamma^{-3}\,t_v^{-1}}$ where $\sigma_{pp}=34.3+1.88\, {\rm S}+0.25\,{\rm S^2}\,\,{\rm mb}$ is the cross section for these interactions with ${\rm S}=\log(E_p/TeV)$ \citep{2006PhRvD..74c4018K}.  Pions $(\pi^{\pm})$ and kaons $(K^{\pm})$ can be produced as products of p-h interactions ({\small$p+h\to X+ \pi^\pm/K^\pm$}). Considering the meson lifetime and p-h interaction time scale 
\be\label{hadr_ts}
t'_{had}\simeq \frac{10\,\pi\,m_p}{\sigma_{pp}}\,  L^{-1}\, \Gamma^{4}\,t_v^{2}\,,
\ee
mesons, then,  in the presence of the magnetic field radiate via synchrotron radiation and also could interact with the primary protons before they decay.
\subsubsection{\bf Pion component}
Equaling the cooling synchrotron radiation ${\small t'_{\pi^+, syn}=\frac{6\pi m^4_{\pi^+}}{\sigma_T\,m_e^2}\,\epsilon^{-1}_B\, L^{-1}\,\Gamma^{2}\,t^2_v\,E'^{-1}_{\pi^+}}$ and  the lifetime of pions ${\small t'_{\pi^+,dec}=\frac{E'_{\pi^+}}{m_{\pi^+}}\,\tau_{\pi^+}}$ with the hadronic time scale (eq. \ref{hadr_ts}), we get  neutrino break energies due to the synchrotron radiation and the lifetime of pions
\be
E_{\nu,\pi^+syn}=1.2\frac{m^4_{\pi^+}\,\sigma_{pp}}{m_p\,\sigma_T\,m_e^2}\epsilon^{-1}_B\,\Gamma\,,
\ee
and
\be
E_{\nu, \pi^+lt}=2.5\frac{\pi\,m_p\,m_{\pi^+}}{\sigma_{pp}}\,\tau^{-1}_{\pi^+}  \,L^{-1}\,\Gamma^{5}\,t^2_v\,,
\ee
respectively. Here $m_{\pi^+}$ is the charged pion mass and $\tau_{\pi^+}$ is the lifetime of the charged pion.
\subsubsection{\bf Kaon component}
Comparing for kaons, the cooling synchrotron radiation ${\small t'_{k^+, syn}=  \frac{6\pi m^4_{k^+}}{\sigma_T\,m_e^2}\,\epsilon^{-1}_B\, L^{-1}\,\Gamma^{2}\,t^2_v \,E'^{-1}_{k^+}}$ and  the lifetime ${\small t'_{k^+,dec}=\frac{E'_{k^+}}{m_{k^+}}\tau_{k^+}}$ with the hadronic time scale (eq. \ref{hadr_ts}), we obtain  neutrino break energies due to the synchrotron radiation and the lifetime of kaons
\be
E_{\nu,k^+syn}=0.3\times \frac{m^4_{k^+}\,\sigma_{pp}}{m_p\,\sigma_T\,m_e^2}\epsilon^{-1}_B\Gamma\,,
\ee
and
\be
E_{\nu, k^+lt}=2.5\frac{\pi\,m_p\,m_{k^+}}{\sigma_{pp}}\,\tau^{-1}_{k^+}  \,L^{-1}\,\Gamma^{5}\,t^2_v\,,
\ee
respectively. Here $m_{k^+}$ is the charged pion mass and $\tau_{k^+}$ is the lifetime of the charged kaon.
%
%
%
%%%%%%%%%%%%%%%%%%%%%%%%%%%%%%%%%%%%%%%%%%%%%%%%%%%%%%%%%%%%
%%%%%%%%%%%%%%%%%%%%%%%%%%%%%%%%%%%%%%%%%%%%%%%%%%%%%%%%%%%%
%%%%%%%%%%%%%%%%%%%%%%%%%%%%%%%%%%%%%%%%%%%%%%%%%%%%%%%%%%%%
%
%
%
\section{Neutrino Expectation}
The expected number of reconstructed neutrino events  in the IceCube experiment can be computed as
\be
N_{ev}=T \,N_A\, \int_{\en^{th}} \sigma_{\nu N}\,M_{eff}\,  \frac{dN_\nu}{d\en}\,d\en\,,
\label{evtrate}
\ee
where $T\simeq$ 4 years is the observation time,   N$_A$=6.022$\times$ 10$^{23}$ g$^{-1}$ is the Avogadro number, $\sigma_{\nu N}=6.78\times 10^{-35}{\rm cm^2}(\en/TeV)^{0.363}$  is  the neutrino-nucleon cross section \citep{1998PhRvD..58i3009G} and $M_{eff}$ is the effective target mass of the IceCube experiment \citep{2013Sci...342E...1I}.  The neutrino spectrum $dN_\nu/d\en$ is computed from p-$\gamma$ and p-h interactions as follows.
\subsection{P-$\gamma$  interactions}
The spectrum of muon neutrino generated by p-$\gamma$ interactions is \citep{2008PhRvD..78c4013K} 
\be\label{neut_pg}
\left(\frac{dN_\nu}{d\en}\right)_{p\gamma}=\frac{V}{4\pi d^2_z}\int_{E_p}\int_{\epsilon} \frac{dN_p}{dE_p}\,  \frac{dN_\epsilon}{d\epsilon}\, F^{p\gamma}_{\nu_\mu}(\eta,x)\frac{dE_p}{E_p}\,d\epsilon\,,
\ee
where $V=\frac43 \pi r_j^3$,  $d_z$ is the luminosity distance from the source and the proton distribution is given by a power law
\be\label{pr_dist}
\frac{dN_p}{dE_p}\simeq A_p\, E_p^{-\alpha}\,,
\ee
with $A_p\simeq n_p/{\rm GeV}$.  The  distribution of target photons  $dN_\epsilon/d\epsilon$ corresponds to the blackbody radiation with temperature and photon density given by eqs. (\ref{enph}) and (\ref{ngamma}), respectively, and the function $F^{p\gamma}_{\nu_\mu}(\eta,x)$ is given in appendix A.
\subsection{P-h interactions}
The spectrum of muon neutrino generated by p-h interactions is  \citep{2006PhRvD..74c4018K}
{\small
\be\label{ph_inter}
\left(\frac{dN_\nu}{d\en}\right)_{ph}=\frac{V\,n_p}{4\pi d^2_z}\int^\infty_{E_p}\sigma_{pp}\, \frac{dN_p}{dE_p}\,F^{pp}_{\nu_\mu} \left(E_p, x\right)\frac{dE_p}{E_p}\,,
\ee
}
where the proton distribution  $dN_p/dE_p$ is given in eq. (\ref{pr_dist}).  The function {\small $F_{\nu_\mu}  = F_{\nu^{(1)}_\mu} + F_{\nu^{(2)}_\mu}$}  comes from the contribution of decay of muon  $F_{\nu^{(1)}_\mu}$ and  from the direct decays of pion $F_{\nu^{(2)}_\mu}$.  Functions $F_{\nu^{(1)}_\mu}$ and $F_{\nu^{(2)}_\mu}$ are explicitly written and explained in appendix B.
\section{Results}
\begin{figure*}
\centering
\includegraphics[width=0.7\textwidth]{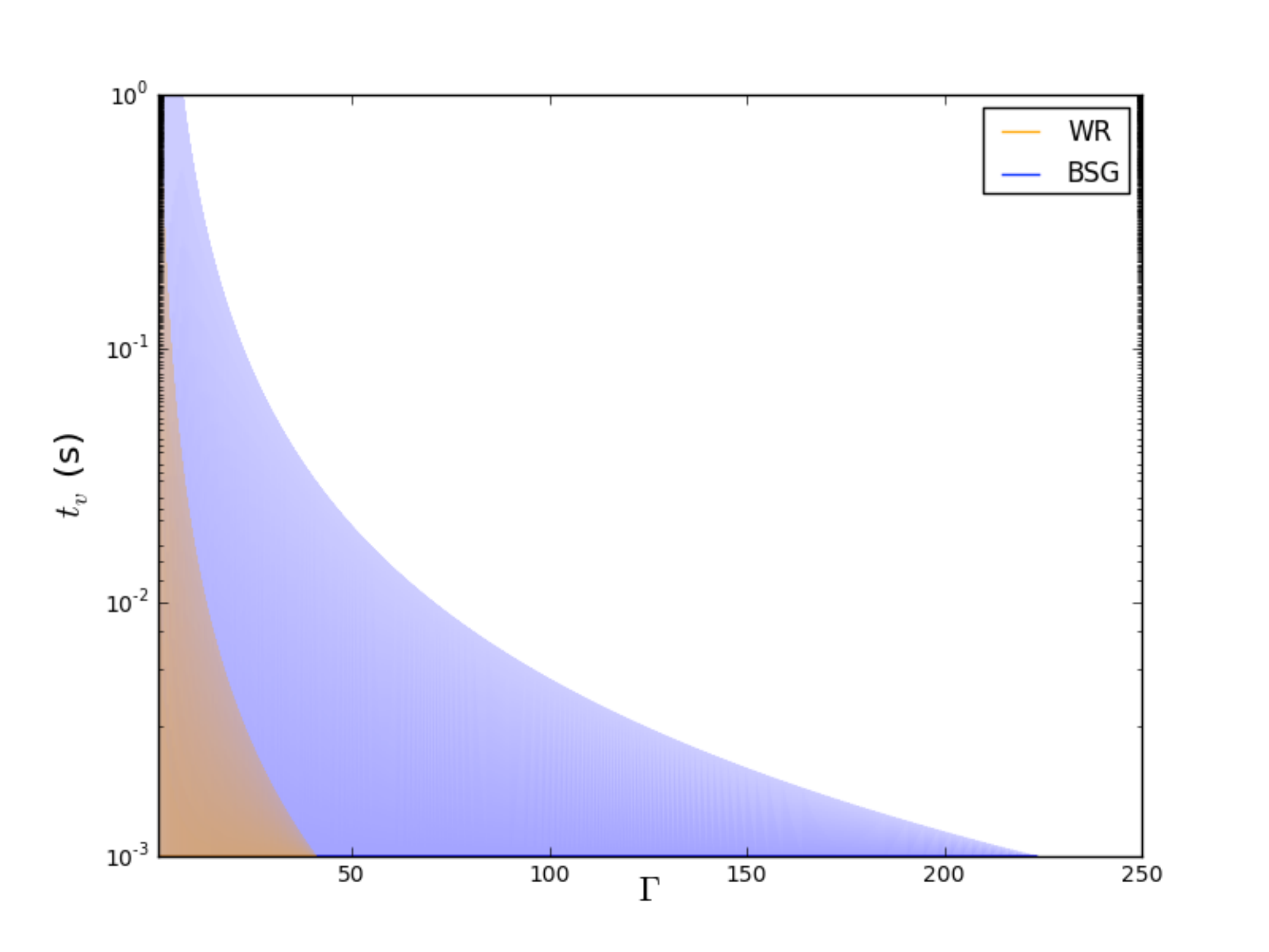}
\caption{Set of values of variability and bulk Lorentz factor for which internal shocks take places  inside a WR (red region) and BSG (blue region). \label{rj}}
\end{figure*}

\begin{figure*}
\centering
\includegraphics[width=0.8\textwidth]{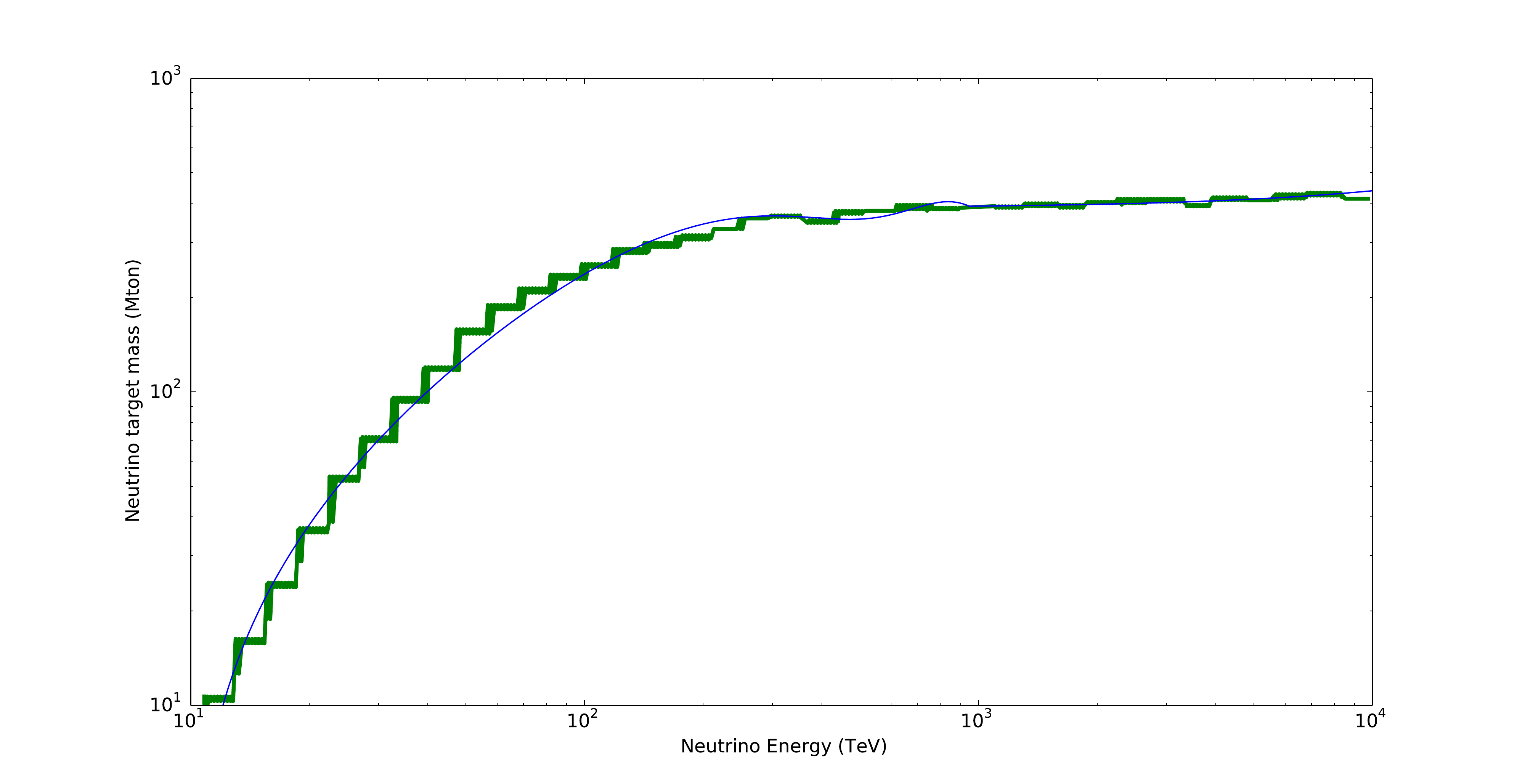}
\caption{Effective target mass of the IceCube experiment as a function of neutrino energy (green line; \citet{2013Sci...342E...1I}). Line in blue color represents the function used to fit the effective target mass.\label{Icecube}}
\end{figure*}

\begin{figure*}
\centering
\includegraphics[width=1.1\textwidth]{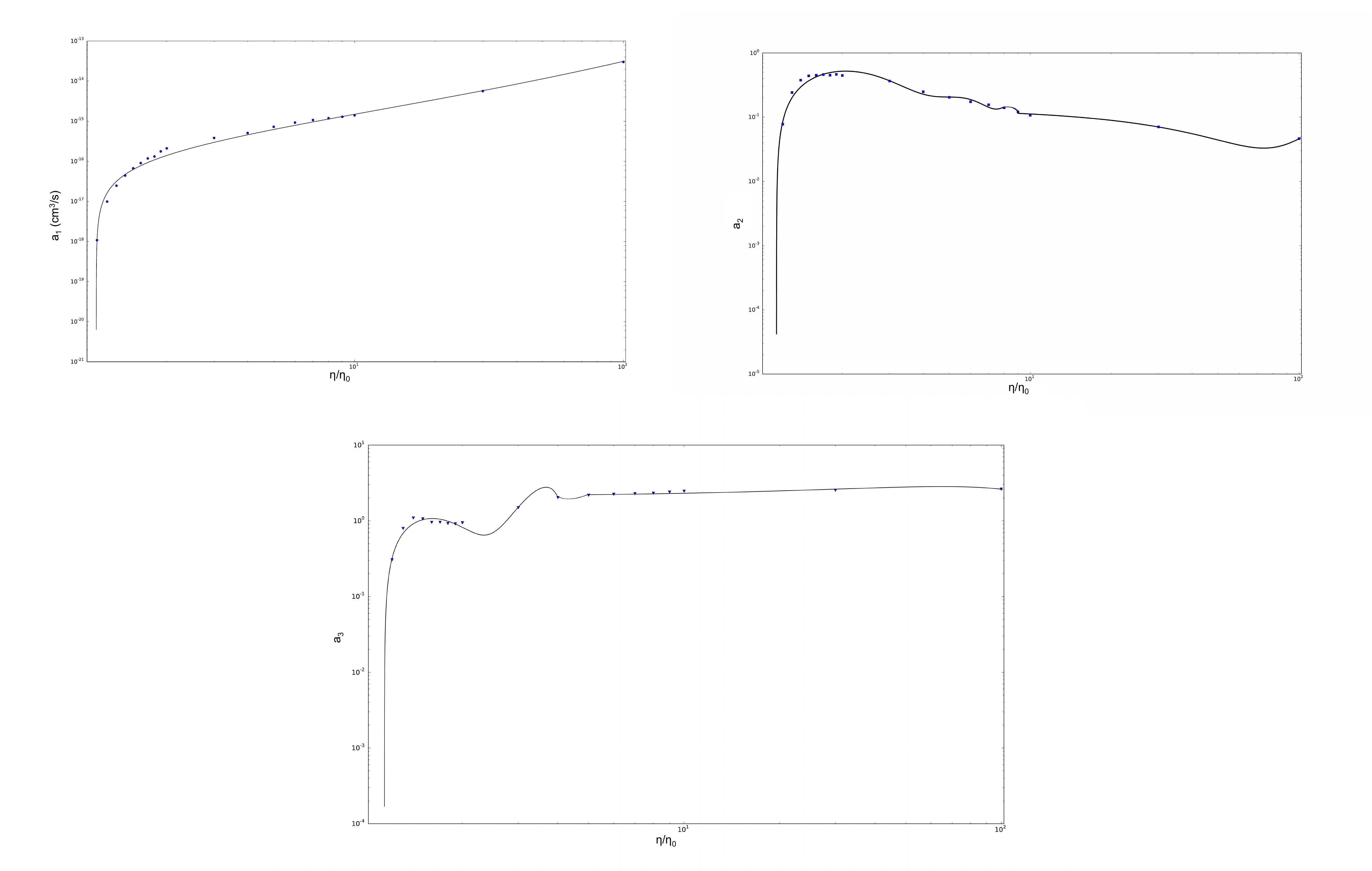}
\caption{Fit of numerical values of parameters $a_1$, $a_2$ and $a_3$ characterizing the neutrino spectrum \citep{2008PhRvD..78c4013K}. \label{parameters}}
\end{figure*}

\begin{figure*}
\centering
\includegraphics[width=1.1\textwidth]{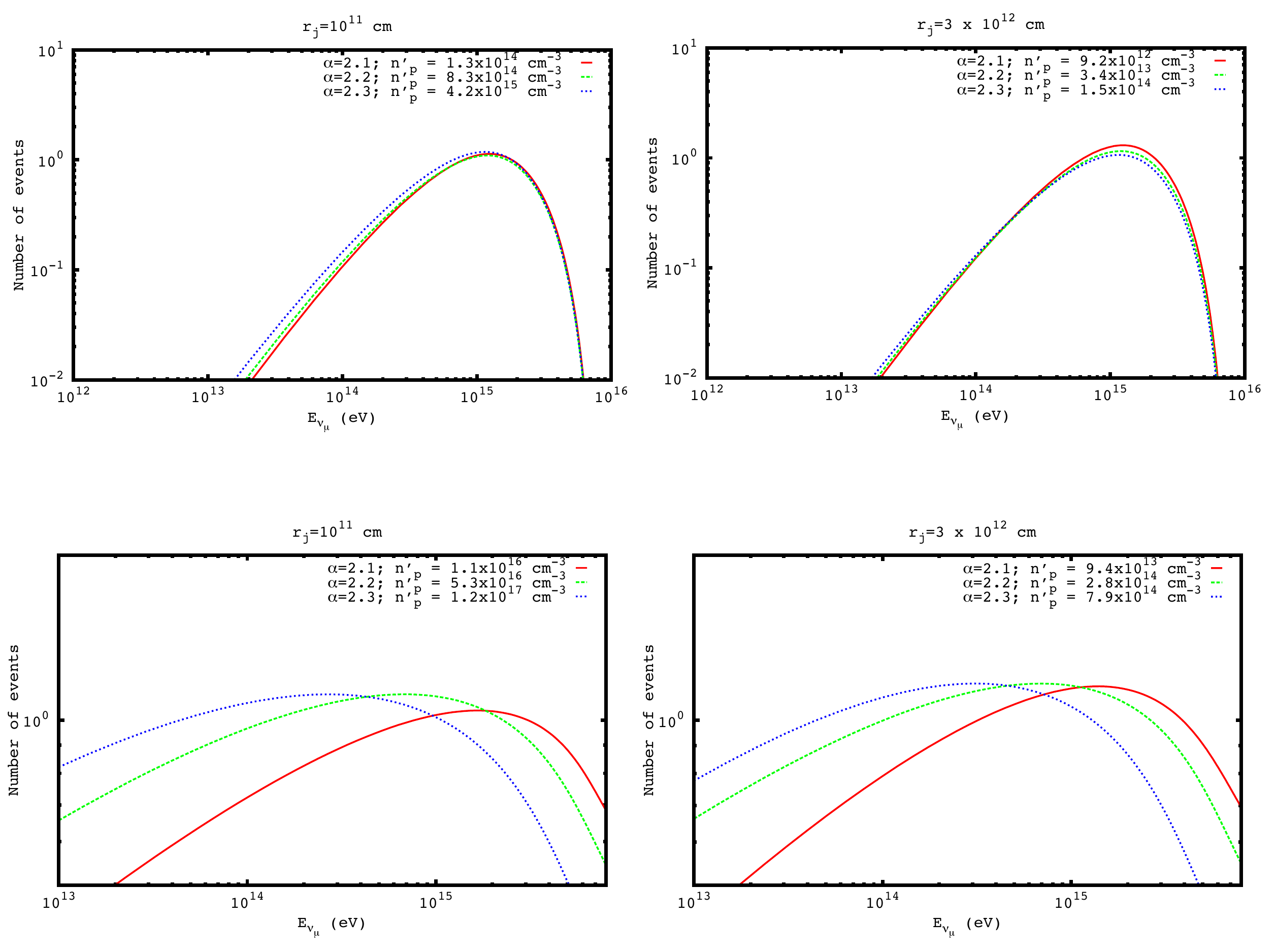}
\caption{Number of events as a function of neutrino energy expected in the IceCube experiment  from the p-$\gamma$ (above panels) and p-h (below panels) interactions. \label{net_expect}}
\end{figure*}

In the CCSNe-GRB  connection framework,  we  have considered WR and  bigger progenitors like BSG stars with formation of jets leading to internal shocks  inside of them.  In these shocks, energy is equipartitioned between magnetic fields and acceleration of  particles.   Electrons and protons are expected to be accelerated in these shocks and after, to be cooled down by synchrotron radiation, inverse Compton and hadronic processes (p-$\gamma$ and p-h interactions).  Both thermalized photons produced by electron synchrotron radiation and hadrons at the shocks serve as targets in the neutrino production through hadronic interactions. In this model, we have shown the production channels  of HE neutrinos  as a function of bulk Lorentz factor, variability time scale, total luminosity and microphysical parameters ($\epsilon_B$ and $\epsilon_e$).  A detailed description such as time scales for all processes (electromagnetic and hadronic) is given in \cite{2014MNRAS.437.2187F}.  Additionally, we have computed the number of events expected in the IceCube experiment. \\
In Figure \ref{rj} is shown the regions for which internal shocks occur inside a WR (red region) and BSG  (blue region) star.  By considering the typical values of variability $10^{-3}\, \leq t_\nu\leq 1\,{\rm s}$, we have found that the bulk Lorentz factor lies in the range $1 \leq \Gamma \leq 40$ for a WR and $1 \leq \Gamma\leq 220$ for a BSG.\\
From eqs. (\ref{wr}) and (\ref{bsg}), and using the typical values of $\theta_0$ and $L$ \citep{2013ApJ...777..162M}, the cylindrical and head radii for a WR and BSG can be written as
{\small
\bary\label{rc_ob}
r_c \simeq
\cases{ 
1.01\times 10^8\, {\rm cm}\,\,t^{1.08}_{1}\,L^{0.81}_{51}\theta^{3.78}_{0,-1}\,M^{-0.81}_{\star,20\,M_\odot}\,R^{0.73}_{\star,11} \,\,\,{\rm for\, WR} \, \cr
3.38\times 10^{9}\, {\rm cm}\,\,t^{2/5}_3\,L^{3/10}_{49.8}\,\theta^{7/5}_{0,-1}\,\hspace{2.1cm}\,{\rm for\, BSG}\,, 
 } 
\eary
}
and
{\small
\bary\label{rh_ob}
r_h \simeq
\cases{ 
9.67\times 10^{10}\, {\rm cm}\,\,t^{1.03}_{1}\,L^{0.35}_{51}\theta^{-0.69}_{0,-1}\,M^{-0.35}_{\star,20\,M_\odot}\,R^{0.31}_{\star,11}  \,\,\,{\rm for\, WR} \, \cr
2.29\times 10^{12}\, {\rm cm}\,\,t^{2/5}_3\,L^{3/10}_{49.8}\,\theta^{7/5}_{0,-1}\,\hspace{2.3cm}\,{\rm for\, BSG}\,. 
 } 
\eary
}
Here we have used $\alpha_\delta$= 2.1.  Taking into account the condition $r_c< r_j$ and eq. (\ref{opt_depth}), we plot contour lines of total luminosity and internal shock radius as a function of bulk Lorentz factor 
for which the  necessary condition for proton acceleration is established in a WR (left panel) and BSG (right panel), as shown in Figure \ref{tau}. From the left panel (WR), one can see that the total luminosity lies in the range $10^{46.5}\leq L\leq 10^{50.5}$ erg/s for an internal shock radius in the range $10^8\leq r_j\leq 10^{11}$ cm and from the right panel (BSG), the total luminosity lies in range  $10^{46}\leq L\leq 10^{49.8}$ erg/s for a radius $10^{10}\leq r_j\leq 10^{12.3}$ cm.   When we consider the maximum values of bulk Lorentz factors $\Gamma \sim 40$ and $\Gamma \sim 220$ for a WR and SGB (see fig. \ref{rj}), respectively,  from eqs. (\ref{opt_depth}) and (\ref{np}), the optical depths become
{\small
\bary\label{opt_med}
\tau \simeq
\cases{ 
2.11\,L_{(50.5 - 48)}r_{j,(8.3 -10.8)}\, \Gamma^{-1}_{1.53}\,\,{\rm for\, WR} \, \cr
2.95\,L_{(49.8 - 48)}\,r_{j,(10.2 - 12.2)}\, \Gamma^{-1}_{2.35}\,\,{\rm for\, BSG}\, 
 } 
\lesssim{\rm min}[\Gamma^2_{rel,1}\,, C^{-1}\Gamma^3_{rel,1} ]\,.
\eary
}
Comparing eqs. (\ref{rc_ob}) and (\ref{opt_med}), it is trivial to demonstrate that accelerated protons could be expected when $r_c < r_j$ and total luminosity in the range $10^{48}<L<10^{50.5}$ erg/s ($10^{48}<L<10^{50}$ erg/s) for a WR (BSG). For instance,  protons accelerated on the surface of either a WR or BSG may be expected under the condition that the jet  would have a luminosity  $L_j\lesssim 10^{48} {\rm erg/s}$ for both a WR and BSG.  Otherwise, under the condition $r_j < r_c$ protons cannot be expected inside the progenitors unless $L>10^{51}$ erg/s or/and $t_v <10^{-3}$ s and $\Gamma > 40$ ($\Gamma > 220$)  for a WR (BSG).    Comparing  the acceleration and the synchrotron time scales, then the maximum energy achieved by accelerated protons in internal shocks  with $r_c < r_j$ and $\tau\sim$ 1 is given by eq. (\ref{Epmax}). \\
%
%%%%%%%%%%%%%%%%%%%%%%%%%%%%%%%%%%%%%%%%%%%%%%%%%%%%%%%%%%%%%%%%%%%%%%%%%%%%%%%%%%%%%%%%%%%%%%%%%%%%%%%%%%%%%%
%%%%%%%%%%%%%%%%%%%%%%%%%%%%%%%%%%%%%%                   pgamma interaction                 %%%%%%%%%%%%%%%%%%%%%%%%%%%%%%%%%%%%%%%%%%%%%%%%%%
%%%%%%%%%%%%%%%%%%%%%%%%%%%%%%%%%%%%%%%%%%%%%%%%%%%%%%%%%%%%%%%%%%%%%%%%%%%%%%%%%%%%%%%%%%%%%%%%%%%%%%%%%%%%%%
%
Figure \ref{Enu_pg} shows neutrino energy (above 30 TeV)  generated by p-$\gamma$ interactions at internal shocks inside of a WR (left; $n_\gamma=10^{16}\, {\rm cm^{-3}}$) and  BSG (right; $n_\gamma=10^{13}\, {\rm cm^{-3}}$) for a luminosity in the range $10^{46}\,\leq\, L \,\leq10^{52}$ erg/s and four values of energy fraction going to accelerate electrons ($\epsilon_e=$0.5,  $5\times 10^{-2}$, $5\times 10^{-3}$ and $5\times 10^{-4}$). Left panels exhibit that the maximum neutrino energy for $L=10^{52}$ erg/s  and $\epsilon_e= 0.5(5\times 10^{-4})$ is $E_\nu=41.7 (9.8)\times 10^{15}$ eV whereas  the neutrino energy is $E_\nu=5.3 (30.1)\times 10^{16}$ for  $L=10^{46}$ erg/s and the same values considered of $\epsilon_e$.  As shown in the right panels, the maximum neutrino energy is $E_\nu=1.4 (8.1)\times 10^{14}$ eV for $L=10^{52}$ erg/s and $\epsilon_e= 0.5(5\times 10^{-4})$, and  $E_\nu=4.2 (25.3)\times 10^{15}$ eV for $L=10^{46}$ erg/s. In panels can be observed a dashed line with $E_\nu$=2.6 PeV.  This value of neutrino event can be obtained for $L<7.3\times10^{46}$ ($7.1\times10^{49}$) erg/s and $\epsilon_e < 0.5$ ($5\times 10^{-4}$) when considered a WR, whereas it is created for $L<10^{51}$ erg/s and $\epsilon_e<0.5$ in a BSG. Finally, based on the condition required for the proton ray acceleration (eq. \ref{opt_med}), we have highlighted the region where parameters are restricted, as indicated in the plot labels.\\
%
%%%%%%%%%%%%%%%%%%%%%%%%%%%%%%%%%%%%%%%%%%%%%%%%%%%%%%%%%%%%%%%%%%%%%%%%%%%%%%%%%%%%%%%%%%%%%%%%%%%%%%%%%%%%%%
%%%%%%%%%%%%%%%%%%%%%%%%%%%%%%%%%%%%%%                   pp interaction                 %%%%%%%%%%%%%%%%%%%%%%%%%%%%%%%%%%%%%%%%%%%%%%%%%%
%%%%%%%%%%%%%%%%%%%%%%%%%%%%%%%%%%%%%%%%%%%%%%%%%%%%%%%%%%%%%%%%%%%%%%%%%%%%%%%%%%%%%%%%%%%%%%%%%%%%%%%%%%%%%%
%
Figure \ref{Enu_pp} shows neutrino energy (above 30 TeV) generated as charged kaon and pion decays in p-h interactions for $10^{-4}\,\leq\epsilon_B\leq\,10^{-1}$ (left) and $10^{46}\,\leq\, L \,\leq10^{52}$ erg/s (right).  Left panel shows that the maximum neutrino energy is  $E_\nu< 3\times 10^{13}$ ($2.1\times 10^{16}$) eV for $\epsilon_B=10^{-1}$ and  $E_\nu=7.1\times 10^{14}$ ($>10^{19}$) eV for $\epsilon_B=10^{-4}$.  Right panel exhibits that  the maximum neutrino energy is $E_\nu=1.1.\times 10^{17}$ ($4.5 \times 10^{21}$) eV for $L=10^{45}$ erg/s  and  $E_\nu< 3\times 10^{13}$ ($5.8 \times 10^{14}$) eV for $L=10^{52}$ erg/s  when neutrino is coming from pion (kaon) decay product inside WR (BSG).  Putting a dashed line to mark the neutrino energy $E_\nu$=2.6 PeV, we can observe that for $\epsilon_{B}<10^{-3}$ ($5.1\times10^{-3}$) neutrinos at this energy could be generated from decay product of pions (kaons)  in a WR whereas for $6.1\times10^{-3}\leq \epsilon_B \leq 0.1$ it could be created in a BSG. The right panel shows that depending on total luminosity, neutrinos coming from a WR and/or BSG could be created as pion and kaon decay products. In this panel, we show  the region where parameters are restricted as showed in the plot label in accordance with the condition required for the proton acceleration (eq. \ref{opt_med}). \\
%
%%%%%%%%%%%%%%%%%%%%%%%%%%%%%%%%%%%%%%%%%%%%%%%%%%%%%%%%%%%%%%%%%%%%%%%%%%%%%%%%%%%%%%%%%%%%%%%%%%%%%%%%%%%%%%
%%%%%%%%%%%%%%%%%%%%%%%%%%%%%%%%%%%%%%                 IceCube detection                %%%%%%%%%%%%%%%%%%%%%%%%%%%%%%%%%%%%%%%%%%%%%%%%%%
%%%%%%%%%%%%%%%%%%%%%%%%%%%%%%%%%%%%%%%%%%%%%%%%%%%%%%%%%%%%%%%%%%%%%%%%%%%%%%%%%%%%%%%%%%%%%%%%%%%%%%%%%%%%%%
%
Requiring the effective target mass of the IceCube experiment \citep{2013Sci...342E...1I}, we use the method of Chi-square $ \chi^2$ minimization as implemented in the ROOT software package \citep{1997NIMPA.389...81B} to fit  the effective target mass as a function of neutrino energy.  Then, the best function that describes the effective target mass  is 
{\small
\bary\label{Veff}
M_{eff} =
\cases{ 
 f_6(E_\nu)\,\,, & $1\, {\rm TeV} < E_\nu < 950\, {\rm TeV}$ \nonumber \cr
 5.19\times 10^{-3}\left(\frac{E_\nu}{TeV}\right)+3.86\times 10^2\,,& $950\, {\rm TeV} < E_\nu < 10^4\, {\rm TeV}$\,,
 } 
\eary
}
where
{\small
\bary\nonumber
f_6(E_\nu)=2.46\times 10^{-15} \left(\frac{E_\nu}{TeV}\right)^6 -1.99\times 10^{-12} \left(\frac{E_\nu}{TeV}\right)^5\cr
-1.09\times 10^{-8} \left(\frac{E_\nu}{TeV}\right)^4\, +2.07\times 10^{-5} \left(\frac{E_\nu}{TeV}\right)^3\cr
 -1.38\times 10^{-2} \left(\frac{E_\nu}{TeV}\right)^2 +3.91 \left(\frac{E_\nu}{TeV}\right) -35.3\,.
\eary
}
It is worth noting that the muon neutrino of the effective target mass was used. Figure \ref{Icecube} shows the effective target mass of the IceCube experiment as a function of neutrino energy. \\
With the p-$\gamma$ spectrum (eq. \ref{neut_pg}) and using the method of Chi-square $ \chi^2$ minimization \citep{1997NIMPA.389...81B}, we have obtained the functions (eqs. \ref{a1}, \ref{a2} and \ref{a3}) that describe the parameters $a_1$, $a_2$ and $a_3$, respectively, as shown in Figure \ref{parameters} and Appendix A.\\
Taking into account the target mass of the IceCube experiment (eq. \ref{Veff}), the p-$\gamma$ (eq. \ref{neut_pg}) and p-hadron (eq. \ref{ph_inter}) spectra, from eq. (\ref{evtrate}) we have computed the number of neutrinos expected  in the IceCube detector. Figure \ref{net_expect} shows the number of neutrinos as a function of energy expected in the IceCube experiment  from the p-$\gamma$ (above panels) and p-h (below panels) interactions for a source located at 10 Mpc.  From p-$\gamma$ interactions we have used a target photon density $n_\gamma\sim 10^{13} {\rm cm^{-3}}$ whereas from p-h interactions a proton density target (as indicated in the label plots) was required.  We compute the upper limit on the proton density in order to obtain one neutrino event at 2.6 PeV for typical values of the spectrum index of proton distribution (2.1$\leq\,\alpha\leq$ 2.3).
%as indicated in the labels,
%
\section{Conclusions}
\begin{figure*}
\centering
\includegraphics[width=1\textwidth]{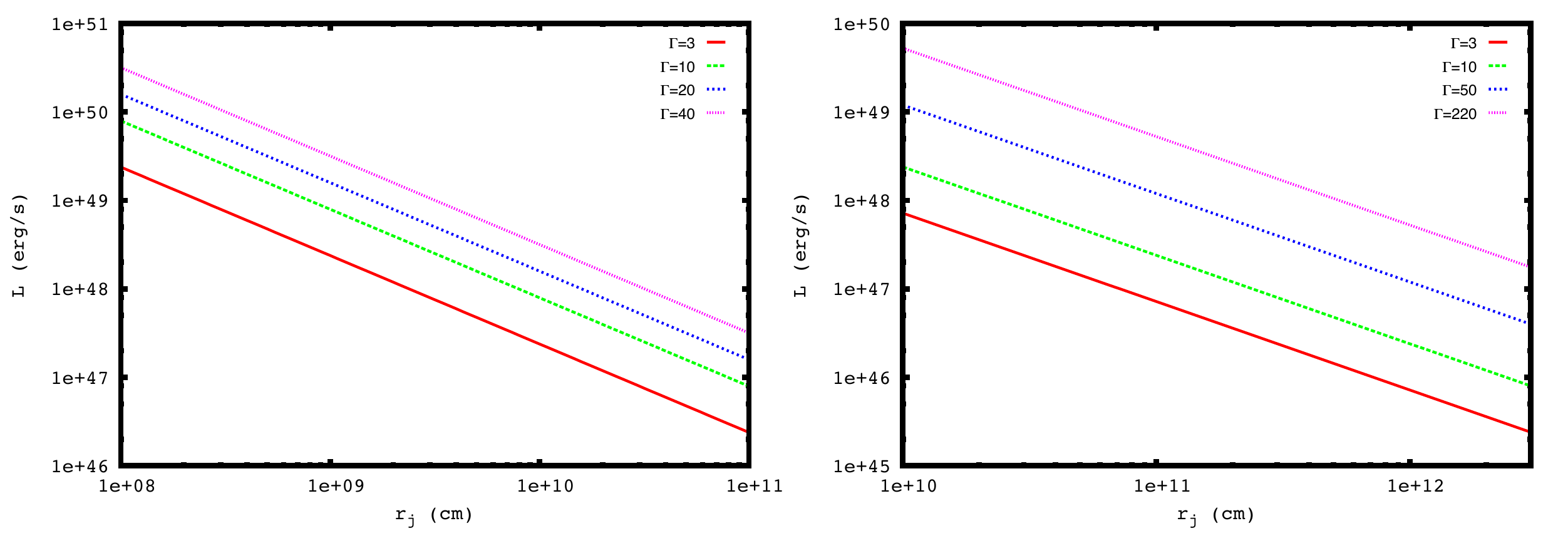}
\caption{Necessary condition for proton acceleration (eq. \ref{opt_depth}) when we consider a WR (left) and BSG (right). \label{tau}}
\end{figure*}

\begin{figure*}
\centering
\includegraphics[width=\textwidth]{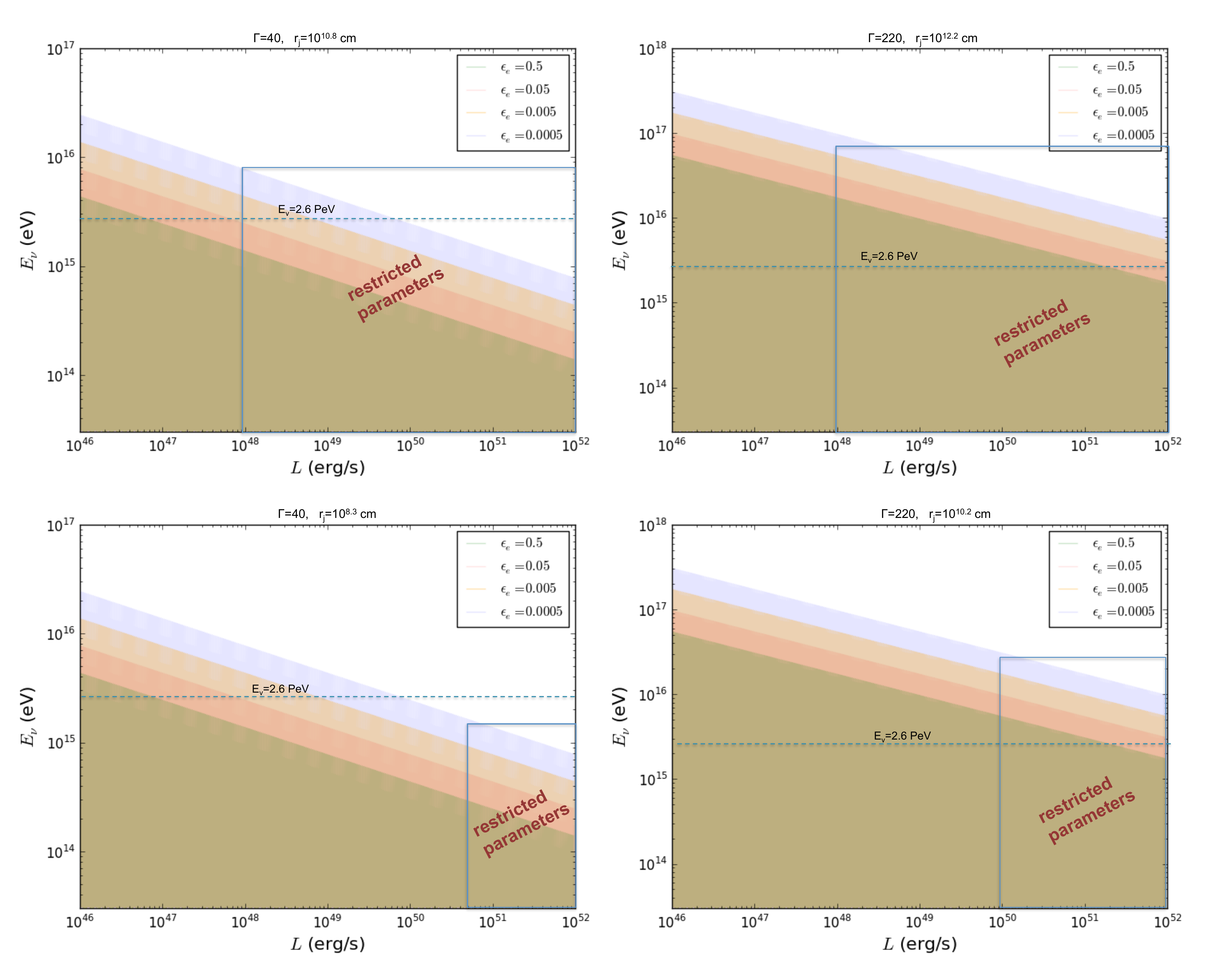}
\caption{Neutrino energy as a function of luminosity generated by p-$\gamma$ interactions at internal shocks inside of a  WR (left) and  BSGs (right). \label{Enu_pg}}
\end{figure*}

\begin{figure*}
\centering
\includegraphics[width=\textwidth]{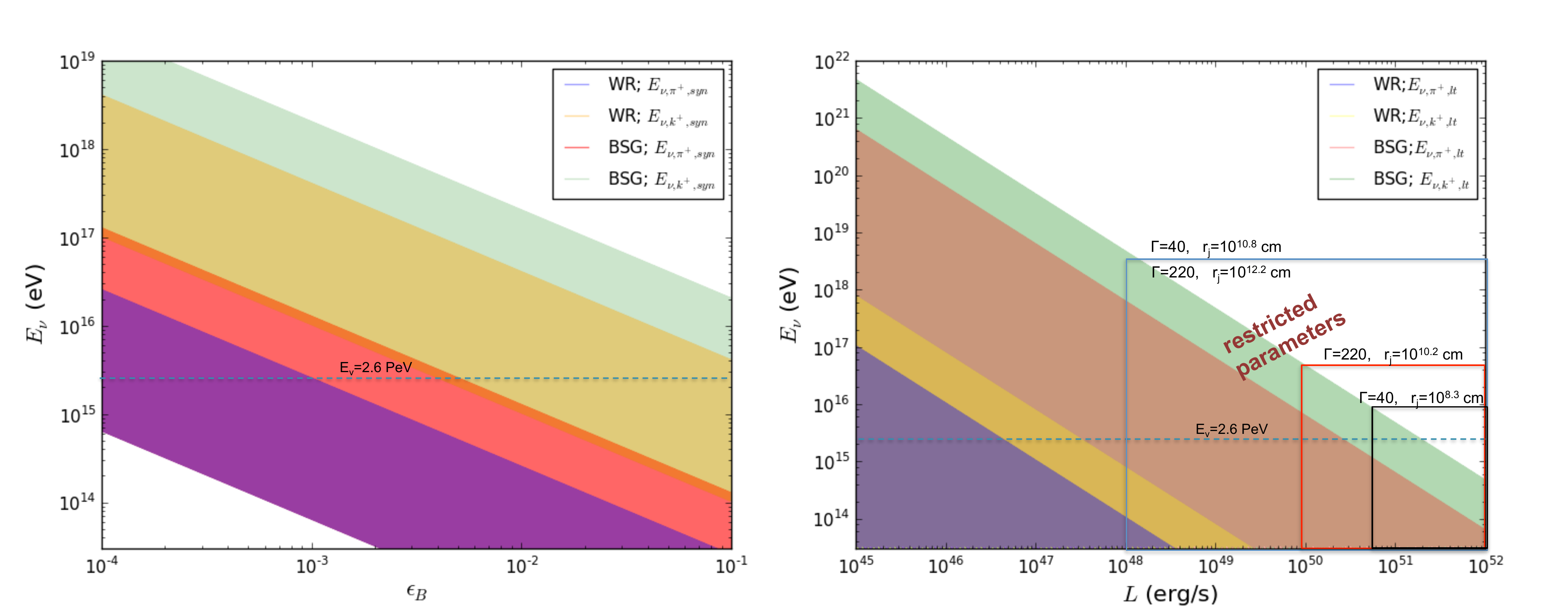}
\caption{Neutrino energy  generated by p-h interactions at internal shocks. synchrotron (left) and  lifetime (right). \label{Enu_pp}}
\end{figure*}

In order to explain the multi-PeV neutrino, we have evoked p-$\gamma$ and p-h interaction model at internal shocks inside the lGRB progenitor stars. We have explored the set of parameter values: $\Gamma$, $t_\nu$, $L$, $\alpha$, $\epsilon_B$ and $\epsilon_e$ to explain the neutrino event at 2.6 PeV. \\
We have computed the characteristic radius and the optical depth to which the shocked jet becomes cylindrical and the condition for proton acceleration inside the star is satisfied.  We have found that protons can hardly be accelerated at internal shocks formed at the distance less than the cylindrical radius $r_j < r_c$ unless $L>10^{51}$ erg/s or $t_v <10^{-3}$ s and $\Gamma > 40$ ($\Gamma > 220$)  for a WR (BSG). Otherwise, protons at internal shocks under the condition $r_c < r_j$ can be accelerated, thus  interacting with hadrons and photons at this place.    We have found that this neutrino could be created by p-$\gamma$ and p-h interactions with a total luminosity and internal shock radius  in the range $10^{46.5}\leq L\leq 10^{50.5}$ erg/s ($10^{46}\leq L\leq 10^{49.8}$ erg/s) and  $10^8\leq r_j\leq 10^{11}$ cm ($10^{10}\leq r_j\leq 10^{12.3}$ cm), respectively, for a WR (BSG) with $\Gamma\leq40$ ($\Gamma\leq220$).  Additionally, the values of equipartition parameters lie in the ranges $\epsilon_e<0.05$ and $10^{-4}\leq \epsilon_B\leq10^{-1}$ which could be a smaller range depending on the progenitor considered.\\
Recently, \cite{2013PhRvL.111l1102M} studied GeV - PeV neutrino production in collimated jets inside progenitors of GRBs, considering both collimation and internal shocks. They showed that although internal shocks in the collimated jet are less favorable for the cosmic ray acceleration than the collimation shocks, due to the meson radiative cooling HE neutrinos around 1 PeV can be only expected from internal shocks. Otherwise,  HE neutrinos from the collimation shocks are expected around 10 TeV due to the strong meson cooling.  They also found that unless $\Gamma \gtrsim10^3$, cosmic rays and HE neutrinos are expected for high-power jets inside a WR.  Authors showed that  llGRBs can be consistent with the astrophysical neutrino background reported by IceCube Collaboration. In this work, we have explored the values of total luminosity, internal shock radius, bulk Lorentz factor and the microphysical parameters so that CCSNe-GRB located at 10 Mpc \citep{and05} could create a 2.6-PeV neutrino event at internal shocks inside the star.\\  
%We show that a multi-PeV neutrino event could be associated to lGRBs with $L\sim 10^{50.5}$ ($\sim 10^{50}$) erg/s  provided that the internal shocks occur at $\sim 10^9$ ($\sim 10^{10.2}$) cm for a WR (BSG).\\
%
In hlGRBs copious target photons  are expected for photo-hadronic interactions, making these GRBs  promising candidates for neutrino detection. Searches for HE neutrinos in spatial and temporal coincidence around this population has been performed, although no neutrinos have been observed.  The null neutrino result reported by IceCube Collaboration in some GRBs \citep{2012Natur.484..351A} could be explained based on the results obtained and showed in figs. (\ref{tau}), (\ref{Enu_pg}) and (\ref{Enu_pp}). We show that a multi-PeV neutrino event could be associated to lGRBs with $L\sim 10^{50.5}$ ($\sim 10^{50}$) erg/s  provided that the internal shocks occur at $\sim 10^9$ ($\sim 10^{10.2}$) cm for a WR (BSG). Otherwise, this PeV neutrino is not expected from hlGRBs with $\Gamma\ll 40$ ($\Gamma\ll 220$) and/or internal shocks on the surface of WR (BSG).\\
By considering a range of GeV - TeV neutrino energies with a flux ratio of $\dot{N}_{\nu_\mu}\simeq\dot{N}_{\bar{\nu}_\mu}\simeq 2 \dot{N}_{\nu_e}\simeq2\dot{N}_{\bar{\nu}_e}$, \cite{2015MNRAS.450.2784F} studied the effect of thermal and magnetized plasma generated in internal shocks and the envelop of star on the neutrino oscillations. The author showed that the resonance lengths lie in the range $l_{res}\sim$ (10$^{12}$ to $10^{14}$) cm, hence depending on the progenitor associated (e.g. WR or SBG) and oscillation parameters, neutrinos would leave the internal shock region and the star in a different flavor to the original. \cite{2015MNRAS.450.2784F}  found that although neutrinos from a few GeV to TeV can oscillate resonantly from one flavor to another, neutrino $>$ PeV can hardly oscillate. Finally, he showed that  a non-significant deviation of the standard flavor ratio (1:1:1) could be expected on Earth.\\
The number of sources with internal shocks inside the stars  may be much larger than those exhibiting one.  Within 10 Mpc, the rate of core-collapse supernovae is $\sim$1 - 3 yr$^{-1}$, with a large contribution of galaxies around  3 - 4 Mpc.   At larger distances,  the expected number of neutrino events in the IceCube detector is still several, and the supernova rate could be $\geq$ 10 yr$^{-1}$ at 20 Mpc \citep{and05}.   Interference effects in the detector by atmospheric neutrino oscillation are very  small (less than 10 \%) due to the short path traveled by neutrinos in comparison with cosmological distances \citep{men07}.
\acknowledgments

We thank C. Dermer, A. M. Sodelberg, B. Zhang,  I. Taboada, K. Murase, W. H. Lee and F. de Colle for useful discussions, Anita M$\ddot{\rm u}$cke and Antonio Galv\'an for helping us to use the SOPHIA program and the TOPCAT team for the useful sky-map tools.\\
%  
%\bibliography{Bib_osc}
%\addcontentsline{toc}{chapter}{Bibliography}

%---- LA BIBLIOGRAPHIE (LISTE PUBLICATIONS EN FIN)----
%\bibliographystyle{elsarticle-num} % (uses file "plain.bst")
%\bibliographystyle{plain} % (uses file "plain.bst")
%\phantomsection
%\addcontentsline{toc}{chapter}{Bibliography}
%\bibliography{Bib_osc}
%\mbox{}
%
%
\appendix
\section{A. Function $F^{p\gamma}_{\nu_\mu}$}
The function $F^{p\gamma}(\eta,x)$ given in eq. (\ref{neut_pg}) is \citep{2008PhRvD..78c4013K} 
\be
F^{p\gamma}_{\nu_\mu}(\eta,x)=a_1\exp\left\{-a_2 \left[\log \left(\frac{x}{x_1} \right) \right]^{a_3}  \right\}\left[\log\left(\frac{2}{1+y^2}   \right)  \right]^\psi \,,
\ee
where {\small $\psi=2.5+1.4\log\left(\frac{\eta}{\eta_0}\right)$} with  {\small $\eta=\frac{4\epsilon\,E_p}{m^2_p}$} and {\small $\eta_0=2\frac{m_\pi}{m_p}+ \frac{m^2_\pi}{m^2_p}\simeq 0.313$}. The variables $y$ and $x$ are 
{\small
\be
y=\frac{x-x_1}{x_2-x_1}\hspace{1cm} {\rm and}\hspace{1cm} x=\frac{\en}{E_p}\hspace{0.5cm}\,,
\ee
}
respectively. The appropriate parameters of the distribution for muon neutrino are: $x_1=0.427\,x_-$ and
{\small
\bary
x_2 =
\cases{ 
0.427\,x_+ \,\,, & $\eta/\eta_0<$2.14  \nonumber \cr
[0.427+0.00729(\eta/\eta_0-2.14)]\,x_+ \,\,,& $2.14<\eta/\eta_0< 10$    \cr
x_+\,\,, &$\eta/\eta_0<10$\,.
 } 
\eary
}
The values of $x_\pm$  correspond to the maximum and minimum pion energies, {\small  $x_\pm=\frac{1}{2(1+\eta)}[\eta + r^2\pm \sqrt{(\eta-r^2-2r)(\eta-r^2+2r)}]$} with {\small $r=m_\pi/m_p\approx$ 0.146}.\\
In accordance with the values $a_1$, $a_2$ and $a_3$ given in \cite{2008PhRvD..78c4013K}) and using the method of Chi-square $ \chi^2$ minimization \citep{1997NIMPA.389...81B},  the best fit of the numerical values of parameters $a_1$, $a_2$ and $a_3$ are
{\small
\be\label{a1}
a_1=1.63\times 10^{-18}  \left(\frac{\eta}{\eta_0}\right)^2+1.49\times 10^{-16}   \left(\frac{\eta}{\eta_0}\right)-1.65\times 10^{-16} \,,
\ee
}
{\small
\bary\label{a2}
a_2 =
\cases{ 
f_6{\left(\frac{\eta}{\eta_0}\right)} \,\,, & 1$<\eta/\eta_0<9$  \nonumber \cr
1.94\times 10^{-5}\left(\frac{\eta}{\eta_0}\right)^2-2.87\times 10^{-3}\left(\frac{\eta}{\eta_0}\right)+0.14  &$\eta/\eta_0>9$\,.
 } 
\eary
}
with
{\small
\bary
f_6{\left(\frac{\eta}{\eta_0}\right)}=-3.88\times 10^{-4}\left(\frac{\eta}{\eta_0}\right)^6+1.30\times 10^{-2}\left(\frac{\eta}{\eta_0}\right)^5-0.17\left(\frac{\eta}{\eta_0}\right)^4\cr
+1.19\left(\frac{\eta}{\eta_0}\right)^3-4.26\left(\frac{\eta}{\eta_0}\right)^2+7.47\left(\frac{\eta}{\eta_0}\right)-4.46 
\eary
}
and
{\small
\bary\label{a3}
a_3 =
\cases{ 
f_4{\left(\frac{\eta}{\eta_0}\right)} \,\,, & 1$<\eta/\eta_0<$4  \nonumber \cr
-1.74\times 10^{-4}\left(\frac{\eta}{\eta_0}\right)^2+2.25\times 10^{-2}\left(\frac{\eta}{\eta_0}\right)+2.10  &$\eta/\eta_0>4$\,.
 } 
\eary
}
with
{\small
\bary
f_4{\left(\frac{\eta}{\eta_0}\right)}&=&-0.97 \left(\frac{\eta}{\eta_0}\right)^4+9.81\left(\frac{\eta}{\eta_0}\right)^3-3.52\times 10\left(\frac{\eta}{\eta_0}\right)^2\cr
&&\hspace{2.5cm}+53.32 \left(\frac{\eta}{\eta_0}\right)-27.81
\eary
}
respectively.
\section{B. Function $F^{pp}_{\nu_\mu}$}
The function $F^{pp}_{\nu_\mu}$ is split in two parts  \citep{2006PhRvD..74c4018K}: one coming from the pion decay $\pi\to\mu\nu_\mu$ ($F^{pp}_{\nu^{(1)}_\mu}$) and the other comes from decay of muon $\mu\to e\bar{\nu}_e\nu_\mu$ ($F^{pp}_{\nu^{(2)}_\mu}$).
\subsection{B1. $F^{pp}_{\nu^{(1)}_\mu}$ from pion decay}
The spectrum of muonic neutrino generated through de direct decay of pion can be written as 
\bary
F_{\nu^{(1)}_\mu}&=&b_1\frac{\log z}{z}\,\left(\frac{1-z^{b_2}}{1+b_3z^{b_2}(1-z^{b_2})}\right)^4\cr 
&&\hspace{0.1cm}\times\left[\frac{1}{\log z} - \frac{4{b_2} z^{b_2}}{1-z^{b_2}} -\frac{4b_3{b_2}z^{b_2}(1-2z^{b_2})}{1+b_3z^{b_2}(1-z^{b_2})}  \right],
\eary
where the variables $z$ and $x$ are {\small $z=\frac{x}{0.427}$} and {\small $x=\frac{\en}{E_p}$}, respectively. The parameters $b_1$, $b_2$ and $b_3$ are
\be
b_1=1.75+0.204\, {\rm S}+0.010\,{\rm S^2}\,,
\ee
\be
b_2=\frac{1}{1.67+0.111\, {\rm S}+0.0038\,{\rm S^2}}\,,
\ee
and
\be
b_3=1.07-0.086\, {\rm S}+0.002\,{\rm S^2}\,,
\ee
with ${\rm S}=\log(E_p/TeV)$.

%%%%%%%%%%%%%%%%%%%%%%%%%%%%%%%%%%%%%%%%%%%%%%%%%%%%%%%%%%%%%%%%%%%%%%%%%%%%%%%%%%%%%%%%%%%%%%%%%%
%%%%%%%%%%%%%%%%%%%%%%%%%%%%%%%%%%%%%%%%%%%%%%%%%%%%%%%%%%%%%%%%%%%%%%%%%%%%%%%%%%%%%%%%%%%%%%%%%%
%
\subsection{B2. $F^{pp}_{\nu^{(2)}_\mu}$ from muon decay}
The spectrum of muonic neutrino generated through de direct decay of muon can be written as 
\be
F_{\nu^{(2)}_\mu}=c_1\frac{[1+c_3(\log x)^2]^3}{x\,(1+\frac{0.3}{x^{c_2}})}(-\log x)^5
\ee
where the parameters $c_1$, $c_2$ and $c_3$ are
\be
c_1=\frac{1}{69.5+2.65\, {\rm S}+0.3\,{\rm S^2}}\,,
\ee
\be
c_2=\frac{1}{(0.201+0.0062\,{\rm S}+0.00042\,{\rm S^2})^{1/4}}\,,
\ee
and
\be
c_3=\frac{0.279+0.141\, {\rm S}+0.0172\,{\rm S^2}}{0.3+(2.3+S)^2}\,.
\ee
\end{document}